\documentstyle[epsf,aps]{revtex}

\begin{document}
\draft

\title{Entropy and typical properties of Nash equilibria in 
two-player games}

\author{
Johannes Berg \cite{jb} and Martin Weigt \cite{mw}\\
{\it \cite{jb} Institut f{\"u}r Theoretische Physik,
Universit{\"a}t Magdeburg, PF 4120, 39106 Magdeburg, Germany\\
\cite{mw}  Laboratoire de Physique Th{\'e}orique
\thanks{Unit{\'e} Mixte de Recherche du Centre National de la
Recherche Scientifique et de l'Ecole Normale Sup{\'e}rieure}, Ecole
Normale Sup{\'e}rieure, 24 rue Lhomond, 75231 Paris cedex 05, France } }

\maketitle
 
\begin{abstract}
We use techniques from the statistical mechanics of disordered systems
to analyse the properties of Nash equilibria of bimatrix games with
large random payoff matrices. By means of an annealed bound, we
calculate their number and analyse the properties of typical Nash
equilibria, which are exponentially dominant in number. We find
that a randomly chosen equilibrium realizes almost always equal
payoffs to either player. This value and the fraction of strategies
played at an equilibrium point are calculated as a function of the
correlation between the two payoff matrices. The picture is
complemented by the calculation of the properties of Nash equilibria
in pure strategies.
\end{abstract}

Game theory seeks to model problems of strategic decision-making
arising in economics, sociology, or international relations. In the
generic set-up, a number of players choose between different
strategies, the combination of which determines the outcome of the
game specified by the payoff to each player. Contrary to the situation
in ordinary optimization problems, these payoffs are in general
different for different players leading to a competitive situation 
where each player tries to maximize his individual payoff.  One of the
cornerstones of modern economics and game theory is therefore the
concept of a Nash equilibrium \cite{Nash}, see also \cite{WJ}, which
describes a situation where no player can unilaterally improve his
payoff by changing his individual strategy given the other players 
all stick to their strategies. However this concept is thought to
suffer from the serious drawback that in a typical game-theoretical
situation there is a large number of Nash equilibria with different
characteristics but no means of telling which one will be chosen by
the players, as would be required of a predictive theory.

This conceptual problem already shows up in the paradigmatic model of
a bimatrix game between two players $X$ and $Y$ where player $X$
chooses strategy $i \in (1 \ldots N)$ with probability $x_i \geq 0$
and player $Y$ chooses strategy $j \in (1 \ldots N)$ with probability
$y_j \geq 0$. The vectors ${\bf x}=(x_1,...,x_N),\ {\bf
y}=(y_1,...,y_N)$ are called mixed strategies and are constrained to
the $(N-1)$-dimensional simplex by normalization.  For a pair of pure
strategies $(i,j)$ the payoff to player $X$ is given by the
corresponding entry in his payoff matrix $a_{ij}$ whereas the payoff
to player $Y$ is given by $b_{ij}$. The expected payoff to player $X$
is thus given by $\nu_x({\bf x},{\bf y})=\sum_{i,j} x_i a_{ij} y_j$
and analogously for player $Y$. Every player has the intention to
maximize his own payoff. A Nash equilibrium (NE) $({\bf x},{\bf y})$ 
is defined by
\begin{eqnarray}
\label{defNE}
  \nu_x({\bf x},{\bf y}) & =& \mbox{max}_{\bf x'}\ 
  \nu_x({\bf x'},{\bf y})  \nonumber\\
  \nu_y({\bf x},{\bf y}) & =& \mbox{max}_{\bf y'}\ 
  \nu_y({\bf x},{\bf y'})
\end{eqnarray}
since in this situation there exists no mixed strategy 
${\bf x}$ which will increase the expected payoff to $X$ 
given $Y$ does not alter his strategy and vice versa for $Y$. Thus 
no player has an incentive to unilaterally change his strategy. 

Apart from applications in economics, politics, sociology, and
mathematical biology, there exists a wide body of mathematical
literature on bimatrix games concerned with fundamental topics such as
exact bounds for e.g. the number of NE \cite{Keiding} and efficient
algorithms for locating them \cite{Stengel}.  For games even of
moderate size a large number of NE are found, forming a set of
disconnected points. In general the different NE all correspond to
different payoffs to the players.

However many situations of interest are characterized by a large number
of possible strategies and complicated relations between the strategic
choices of the players and the resulting payoffs.  In such cases it is
tempting to model the payoffs by random matrices in order to calculate
{\it typical} properties of the game. This idea is frequently used in
the statistical mechanics approach to complex systems such as spin
glasses \cite{MPV,Young}, neural networks \cite{HKP}, evolutionary
models \cite{DiOp}, or hard optimization problems \cite{TSPMoZe}.
Recently this approach has been used to investigate the typical
properties of so-called zero-sum games \cite{BergEngel}.

In this Letter we investigate the properties of Nash
equilibria in large bimatrix games with random payoffs, {\it i.e.}
with random entries of the payoff matrices. Using
techniques from the statistical mechanics of disordered systems we
estimate the number of NE with a given payoff.  We find that the NE
are exponentially dominated in number by equilibria with a certain
payoff. 

In this approach characteristics of the game are encoded in the
distribution of payoff matrices -- with only a few parameters --
instead of the payoff matrices themselves. Clearly the choice of the
probability distribution to be averaged over can influence the
results. However a number of simplifying observations may be made: the
two payoff matrices may be multiplied by any constant or have any
constant added to them without changing the properties of the game in
any material way. Thus there is no loss of generality involved in
considering payoffs of order $N^{-1/2}$ and of zero mean. We assume
that the entries of the payoff matrices at different sites are
identically and independently distributed.  In the thermodynamic limit
one finds that only the first two moments of the payoff distribution
are relevant so the entries of the payoff matrices may be considered
to be Gaussian distributed. The only property of the distribution of
payoffs which is not fixed by these specifications is the correlation
$\kappa$ between entries of the same site of the two payoff matrices.
We thus choose the entries of the payoff matrices to be drawn randomly
according to the probability distribution
\begin{equation}
\label{eq:payensemble}
 p(\{a_{ij}\},\{b_{ij}\}) = \prod_{ij} \frac{N}{2\pi\sqrt{1-\kappa^2}}
 \exp\left(-\frac{N (a_{ij}^2-2 \kappa a_{ij} b_{ij} + 
 b_{ij}^2 )}{2(1-\kappa^2)} \right),
\end{equation}
{\it i.e.} a Gaussian distribution with zero mean, variance $1/N$ and
correlation $\overline{a_{ij}b_{kl}}=\kappa\delta_{ik}\delta_{jl}/N$
for all pairs $(i,j)$ and $(k,l)$.  Here and in the following, the
overbar $\overline{\ \cdot   \ }$ denotes the average over the payoff 
distribution (\ref{eq:payensemble}).  

Thus for $\kappa=-1$ one finds a Dirac-delta factor
$\delta(a_{ij}+b_{ij})$ in (\ref{eq:payensemble}) corresponding to a
zero-sum game. $\kappa=0$ corresponds to uncorrelated payoff matrices
and $\kappa=1$ gives the so-called symmetric case $a_{ij}=b_{ij}$
where the two players always receive identical payoffs. The parameter
$\kappa$ consequently describes the degree of similarity between the
payoffs to either player and may be used to continuously tune the game
from a zero-sum game to a purely symmetric one.  The former admits no
cooperation at all between the players. Negative $\kappa$ correspond
to an on average competitive situation, whereas for increasing
$\kappa$ there are more and more pairs of strategies which are
beneficial to both players.

As a starting point for the statistical
mechanics calculations, we remark that (\ref{defNE}) is equivalent to
the set of inequalities 
\begin{eqnarray}
\label{eq:NE}
 \sum_j a_{ij} y_j-\nu_x({\bf x},{\bf y}) \leq  0 && \ \ \ \forall i
 \nonumber\\
\sum_i x_i b_{ij}-\nu_y({\bf x},{\bf y}) \leq  0 && \ \ \ \forall j\ .
\end{eqnarray}
Due to the non-negativity of the probabilities $x_i$ and $y_j$ this
directly results in the local equations specifying a NE with payoffs
$\nu_x$ and $\nu_y$
\begin{eqnarray}
\label{eq:NElocal}
 x_i \left(\sum_j a_{ij} y_j -\nu_x \right)=0 &\ \ \ 
 \left(\sum_j a_{ij} y_j -\nu_x \right)\leq 0& \ \ \ \forall i
  \nonumber\\
 y_j \left(\sum_i x_i b_{ij} -\nu_y \right)=0 &\ \ \ 
 \left(\sum_i x_i b_{ij} -\nu_y \right)\leq 0& \ \ \ \forall j \ .
\end{eqnarray}
We introduce real-valued variables ${\bf \tilde{x}}$ and
${\bf \tilde{y}}$ with $x_i=\tilde{x}_i \Theta(\tilde{x}_i)$ (where
$\Theta(x)$ is the Heaviside step-function) and likewise for $y_j$.
Then the last conditions (\ref{eq:NElocal}) may be written as
\begin{eqnarray}
I^x_i(\tilde{{\bf x}},\tilde{{\bf y}}) 
&=\tilde{x}_i \Theta(-\tilde{x}_i) - (\sum_j 
a_{ij}\tilde{y}_j \Theta(\tilde{y}_j)  -\nu_x) =& 0 \nonumber \\
I^y_j(\tilde{{\bf x}},\tilde{{\bf y}}) 
&=\tilde{y}_j \Theta(-\tilde{y}_j) - (\sum_j 
\tilde{x}_i \Theta(\tilde{x}_i) b_{ij} -\nu_y) =& 0 \ ,
\end{eqnarray}
{\it i.e.} we have constructed indicator functions $I^x_i$ and $I^y_i$ 
which are zero at a NE $({\bf x},{\bf y})$ and non-zero elsewhere
\cite{opper}.

The number ${\cal N}(\nu_x,\nu_y)$ of NE with payoffs $\nu_x$ and
 $\nu_y$ may thus be calculated from
\begin{equation}
\label{part}
{\cal N}(\nu_x,\nu_y)=\int d{\bf \tilde{x}}\ d{\bf \tilde{y}}\ 
\delta\left(\sum_i \tilde{x}_i \Theta(\tilde{x}_i) -N\right) \ 
\delta\left(\sum_j \tilde{y}_j \Theta(\tilde{y}_j) -N\right)  
\prod_i \delta\left(I_i^x(\tilde{{\bf x}},\tilde{{\bf y}})\right) 
\prod_j \delta\left(I_j^y(\tilde{{\bf x}},\tilde{{\bf y}})\right) \ 
\left\| \frac {\partial ({\bf I^x},{\bf I^y})}
{\partial ({\bf \tilde{x}},{\bf \tilde{y}})} \right\| \ ,
\end{equation}
where the mixed strategies have been rescaled to $\sum_i x_i=\sum_j
y_j=N$ for convenience.  Although we expect an exponential number of
NE, so $\ln {\cal N}$ is an extensive quantity, we calculate the
so-called annealed average $\ln \overline{ {\cal N} }$ instead of the
quenched one $\overline{\ln {\cal N} }$. This gives an exact upper
bound to the typical number of NE at given payoffs
\footnote{The annealed approximation does not differ by more than 3\%
in the entropy and by 10\% in the order parameters from the results of
the more complicated replica calculation of the quenched average,
which will be presented elsewhere.}.  Using integral representations
of the delta functions in (\ref{part}), averaging over the disorder
(\ref{eq:payensemble}), and introducing the order parameters
$q^x=N^{-1} \sum_i \tilde{x}_i ^2 \Theta(\tilde{x}_i)$,
$R^x=N^{-1} \sum_i i\hat{x}_i \tilde{x}_i \Theta(\tilde{x}_i)$, 
and similarly for player $Y$ we obtain
\begin{eqnarray}
\label{part2}
\overline{ {\cal N}(\nu_x,\nu_y) } &=&
\int \frac{dq^{x,y} d\hat{q}^{x,y}}{(2\pi/N)^2} 
\int \frac{dE^{x,y}}{(2\pi/N)^2}  \int \frac{dR^{x,y}}{2\pi/(N\kappa)}\ 
\exp\{iN (q^{x} \hat{q}^{x} +q^{y} \hat{q}^{y} -\kappa R^x
R^y + E ^x + E ^y)\} \nonumber \\ 
&&\int\prod_i\frac{d\tilde{x}_i d \hat{x}_i}{(2 \pi)} 
  \int\prod_j\frac{d\tilde{y}_j d \hat{y}_j}{(2 \pi)}\ 
\exp \left\{ -\frac{i \hat{q}^x}{2} \sum_i  \tilde{x}_i ^2 
\Theta(\tilde{x}_i) + i \kappa R^y \sum_i i \hat{x}_i \tilde{x}_i 
\Theta(\tilde{x}_i) -\frac{q^y}{2}\sum_i \hat{x}_i ^2 \right.
 \nonumber \\ 
&&\ \ \ \left. -i\sum_i \tilde{x}_i \hat{x}_i \Theta(-x_i) 
  -i \nu_x \sum_i \hat{x}_i-i E ^x \sum_i\tilde{x}_i
  \Theta(x_i) +(x\leftrightarrow y \ \mbox{with}\ iR^y \rightarrow
  R^x)\right\} \| \det(B) \| \ ,
\end{eqnarray}
with $H(x)=\int_x^\infty dx'\ \exp(-x'^2/2)/\sqrt{2\pi}$. Here we have
used the assumption that the normalizing determinant $\| \det(B) \| =
\| \frac {\partial ({\bf I^x},{\bf I^y})}{\partial ({\bf
\tilde{x}},{\bf \tilde{y}})} \|$ is self-averaging and is effectively
uncorrelated with the rest of expression (\ref{part2}) and may
therefore be averaged over the disorder independently of the rest of
the expression \cite{opper} \footnote{This assumption ha been verified by explicitly
including the calculation of the effective normalizing determinant
into the disorder average.}. We obtain
\begin{equation}
\| \det(B) \|=\exp \left\{\frac{N}{2}(p_x\ln p_x -p_x + p_y\ln p_y
-p_y) \right\} \ ,
\end{equation} 
where the fraction of strategies played with non-zero probability
$p_x=N^{-1}\sum_i \Theta(\tilde{x}_i)$ is determined self-consistently
by introducing $p_x$, $p_y$ and their conjugates $\hat{p}_x$,
$\hat{p}_y$ in (\ref{part2}). After the transformation $i\hat{q}^{x,y}
\rightarrow \hat{q}^{x,y}$ and similarly for $E^{x,y},\hat{p}^{x,y},
R^y$, the integrals over ${\bf \tilde{x}} , {\bf \hat{x}} ,{\bf
\tilde{y}} , {\bf \hat{y}}$ may be performed. In the limit of large
payoff matrices $N \to \infty$, the integrals over the order parameters
are dominated by their saddle point. Furthermore (\ref{part2}) is now
symmetric under an interchange of the players, and the maximum (and
hence exponentially in $N$ dominating) number of NE is found at equal
payoffs to each player. Thus it makes sense to restrict the analysis
to the case $\nu_x=\nu_y=\nu$. We obtain
\begin{eqnarray}
\label{entropy}
S_\kappa(\nu)=\frac{1}{N} \ln \overline{ {\cal N}(\nu,\nu) }
&=&\mbox{2 extr}_{q,\hat{q},R,E,p} 
 \left[\frac{q \hat{q}}{2} - \frac{\kappa R^2}{2} -\frac{ p}{2} 
\right. \nonumber \\
&&\left. +\ln \left(H\left(-\frac{\nu}{\sqrt{q}}\right) + 
 \sqrt{\frac{p}{q \hat{q}+\kappa^2 R^2}}
\exp\left\{-\frac{\nu^2}{2q}+\frac{(E-\kappa R \nu/q)^2}{
2\hat{q}+2\kappa^2 R^2/q}\right\} H\left(\frac{E-\kappa R \nu/q}{\sqrt{
\hat{q}+\kappa^2 R^2/q}} \right) \right)\right]
\end{eqnarray} 
From this expression the statistical properties of the entire spectrum
of player-symmetric NE may be deduced. In the thermodynamic limit, a
randomly chosen NE will give the payoff
$\nu=\nu_y=\nu_x=\mbox{argmax}S_\kappa(\nu)$ with probability one,
because the number of NE with this payoff
${\cal N}(\nu,\nu)=\exp\{N\mbox{max}S_\kappa(\nu)\}$ is exponentially 
larger than the number of all other NE. Similarly, the self-overlap
$q=\sum_i x_i^2=\sum_j x_j^2$ and the fraction of strategies played with
non-zero probability $p=p_x=p_y$ will take on their saddle-point
values of (\ref{entropy}) evaluated at the maximum of
$S_{\kappa}(\nu)$ with probability one.

\begin{figure}[htb]
 \epsfysize=6.5cm
      \epsffile{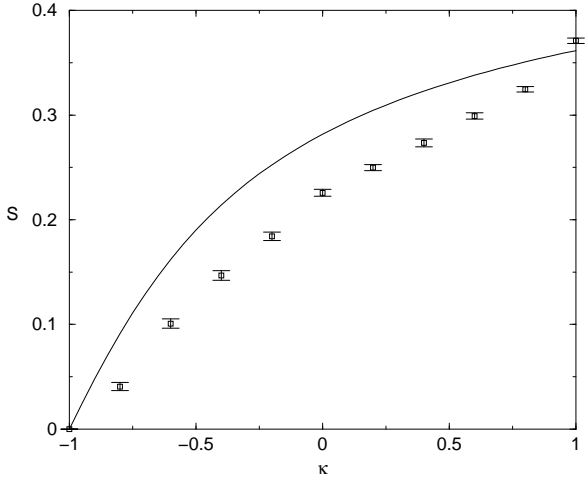}
\caption{
\label{ops_nu}
The entropy $S_{\kappa}$ of the NE which exponentially dominate the
spectrum $S_\kappa(\nu)$. The analytic results are compared with
numerical simulations for $N=18$ averaged over 100 samples.}
\end{figure}

Figure \ref{ops_nu} shows the entropy $S_{\kappa}$ at the maximum of
$S_{\kappa}(\nu)$ as a function of $\kappa$. 
The numerical results were obtained by enumerating all NE for N=18
\cite{Stengel,AvisFukuda}. Given that enumeration is only possible 
for small sample sizes, the agreement between the annealed approximation 
and the numerical results is quite good. 

\begin{figure}[htb]
 \epsfysize=6.5cm
      \epsffile{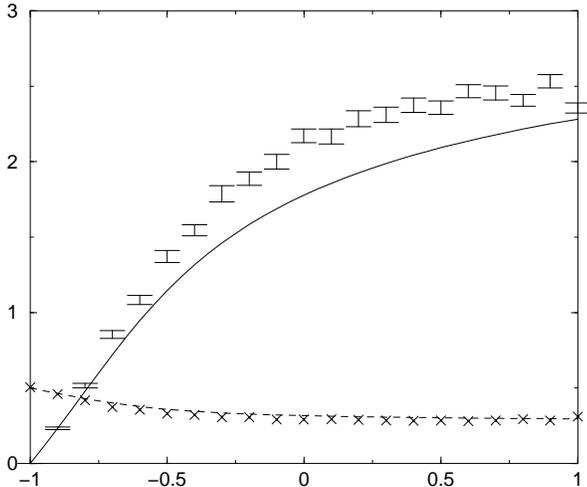}
\caption{
\label{ops_nu2}
The payoff $\nu$ (solid line) and the fraction $p$ of strategies played with 
non-zero probability (dashed line). The analytic results are compared with numerical simulations 
for $N=50$ averaged over 200 samples.
}
\end{figure}

Figure \ref{ops_nu2} shows the payoff $\nu$ and the fraction $p$ 
of strategies played with non-zero probability again as a
function of $\kappa$. The numerical results were obtained for N=50 by
using the iterated Lemke-Howson algorithm \cite{LH,Stengel}, which
locates a single but typical NE. The deviation between analytic and 
numerical results may be accounted for by the use of the 
annealed calculation and finite-size effects. 

For zero-sum games ($\kappa=-1$) there is only a single NE
\cite{BergEngel} so $S_{\kappa=0}=0$, and the payoff is zero due to
the symmetry between the players.  As $\kappa$ is increased the number
of NE rises. The maximum of the typical number of NE is reached for
the case of symmetric games, where $S_{\kappa=1} \sim 0.362$. This
result may be compared with a rigorous upper bound derived using
geometric methods \cite{Keiding,Stengel}, which states that for any
non-degenerate $N$-by-$N$ bimatrix game with large $N$ there are at
most $e^{0.955 N}$ equilibrium points.  Thus the typical-case scenario
investigated here does not saturate this bound.

The increase of $\nu$ with $\kappa$ may be understood as
follows: At increasing $\kappa$ the outcome of a pair of strategies
$(i,j)$ which is beneficial to player $X$ say, tends to become more
beneficial to player $Y$.  As a result players focus on these
strategies and the payoff to both players rises.  By the same token
the fraction $p$ of strategies which are played with non-zero
probability at a NE decreases with $\kappa$ and the self-overlap of
the mixed strategies $q$ increases.

Even though the properties of the NE at a given $\kappa$ are dominated
by the peak of $S_{\kappa}(\nu)$, there is an important qualitative
difference between the curves of different values of $\kappa$. For
$-1<\kappa<\kappa_c \sim -.59$, $S_{\kappa}(\nu)$ reaches its maximum
and then decreases crossing the $S=0$ axis at some point.  For
$\kappa>\kappa_c$ however, $S_{\kappa}(\nu \to \infty) \to 0+$ and $p
\to 0+$.  Hence there is an exponential number of NE offering an
arbitrarily large (but finite) payoff to either player.

This observation can be corroborated by considering the
extreme case of pure strategy Nash equilibria (PSNE) $(i,j)$ where
each player only plays a single (pure) strategy \cite{Goldman},
{\it i.e.} $x_i=N\delta_{i'i},\ y_j=N\delta_{j'j},\ \forall i',j'=1,...,N$.  
According to (\ref{eq:NE}), a PSNE corresponds to a site $(i,j)$ which 
is simultaneously a maximum of the column $a_{i'j}$ and a maximum of 
the corresponding row of $b_{ij'}$.  The average of the number
$\cal{M}$ of PSNE is thus given by
\begin{eqnarray}
\label{nPSNE}
 \overline{\cal{M}} &=& 
 \overline{\sum_{ij} \prod_{k \neq i} \Theta(a_{ij}-a_{kj}) 
                     \prod_{l \neq j} \Theta(b_{ij}-b_{il})} \nonumber \\
 &=& N^2 \int da \, db \, p(a,b) H^{N-1}(-\sqrt{N}a) H^{N-1}(-\sqrt{N}b) 
\end{eqnarray}
where $p(a,b)$ denotes one of the Gaussian factors in 
(\ref{eq:payensemble}). 

\begin{figure}[htb]
 \epsfysize=6.5cm
      \epsffile{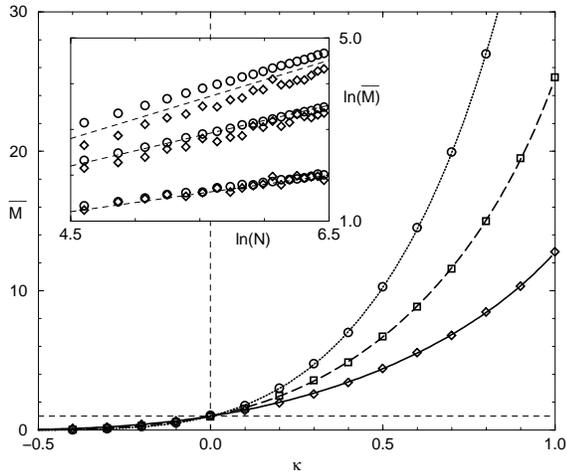}
\caption{Average number $\overline{\cal{M}}$ of PSNE for $N=25,50,100$
(full, dashed, dotted lines). The lines show analytical result (\ref{nPSNE}), 
the symbols numerical results, averaged over 1000 samples. Inset: 
Numerical results for the logarithm of $\overline{\cal{M}}$ 
(circles) and of $\overline{{\cal M}^2}-\overline{\cal{M}}^2$ with
$\kappa=0.25,0.5,0.75$ (bottom to top) against $\ln N$. The dashed
lines show the asymptotic slope $2\kappa/(1+\kappa)$ of 
$\overline{\cal{M}}$.}
\label{fig:pure}
\end{figure}

Figure \ref{fig:pure} shows the average number of PSNE as well as its
variance as a function of $\kappa$. For $N \to \infty$ one finds from
(\ref{nPSNE}) that $\overline{\cal{M}}$ scales with
$N^{2\kappa/(1+\kappa)}$, so the number of PSNE vanishes for
$\kappa<0$, at $\kappa=0$ there is on average $1$ PSNE, and for
$\kappa>0$ the number of PSNE diverges. The increase of the number of
PSNE with $\kappa$ may be understood as follows.  At low $\kappa$, $a$
and $b$ are anticorrelated, so in the extreme case $\kappa=-1$ a PSNE
corresponds to a site $(i,j)$ which is simultaneously a maximum of the
column $a_{i'j}$ and a minimum of the row $a_{ij'}=-b_{ij'}$. In the
thermodynamic limit, such points are exponentially rare. For large
$\kappa$, the condition for a PSNE is much easier to fulfill and in
the case $\kappa=+1$, there is always at least one PSNE, namely the
maximum entry of $a_{ij}=b_{ij}$.

Note that apart from the point $\kappa=0$, the number of PSNE is also 
self-averaging as the relative sample-to-sample fluctuations 
of $\cal{M}$ are found to vanish in the large $N$-limit. Furthermore 
for all $\kappa>-1$ the typical PSNE payoffs to leading 
orders of $N$ are found to be independent of $\kappa$, 
they are $\sqrt{N(2\ln N - \ln\ln N + O(1))}$.

In conclusion we have used methods from the statistical mechanics of
disordered systems to describe the typical properties of Nash
equilibria in bimatrix games with large random payoff matrices. We
find that in the thermodynamic limit for a randomly chosen Nash
equilibrium quantities such as the fraction of strategies played with
non-zero probability, the self-overlap of the mixed strategies and
most importantly the payoff to either player take on a specific value
with probability 1.  We have analytically calculated these quantities
in the annealed approximation as a function of the correlation
$\kappa$ between the payoff matrices and found good agreement with
numerical simulations. Furthermore the properties of Nash equilibria
in pure strategies were calculated as a limiting case.  A number of
extensions of this work may be considered, including the case where
the players have different numbers of strategies at their disposal and
the generalisation to games of several players.

{\bf Acknowledgments}: We gratefully acknowledge crucial discussions 
with L.~Cugliandolo, J.~Kurchan, and especially M.~Opper. Many thanks to 
B.~von~Stengel for help with the numerical simulations.
M.W. acknowledges financial support by the German Academic
Exchange Service (DAAD) and J.B by the Studien\-stiftung des 
Deutschen Volkes. J.B. would like to thank the LPTENS, Paris, where 
some of this work was done, for hospitality. 

\bibliographystyle{unsrt}

\end{document}